\documentclass[aip,journal-option,reprint]{revtex4-2}

\usepackage{natbib,graphicx,dcolumn,bm,amsmath,color,bm,subfigure,soul}

\usepackage{tikz}
\usetikzlibrary{calc}

\newcommand{\eq}[1]{Eq.~(\ref{#1})}

\newcommand{\fig}[1]{Fig.~\ref{#1}}

\definecolor{amber}{rgb}{1.0, 0.75, 0.0}

\newcommand{\red}[1]{ {\color{red} #1}}

\newcommand{\eeq}{ \end{equation} }
\newcommand{\beq}{ \begin{equation} }

\newcommand{\eea}{ \end{align} }
\newcommand{\bhea}{ \begin{align} }

\newcommand{\bhu}{{\bf \hat{u}}}

\newcommand{\bhe}{{\bf \hat{e}}}

\newcommand{\bhv}{{\bf \hat{v}}}
\newcommand{\bhw}{{\bf \hat{w}}}

\newcommand{\br}{ {\bf p }}

\newcommand{\bn}{ {\bf \hat{n} }}

\begin{document}

\title{Shape enantiomerism in semi-rigid polymer liquid crystals}

\author{S. Biswas}
\affiliation{Laboratoire de Physique des Solides - UMR 8502, CNRS,  Universit\'{e} Paris-Saclay, 91405 Orsay, France}
\affiliation{Institut Charles Sadron, Université de Strasbourg \& CNRS, 23 rue du Loess, 67034 Strasbourg Cedex, France}

\author{W. S. Fall}
\email{william.fall@cnrs.fr}
\affiliation{Institut Charles Sadron, Université de Strasbourg \& CNRS, 23 rue du Loess, 67034 Strasbourg Cedex, France}

\author{H. H. Wensink}
\email{rik.wensink@cnrs.fr}
\affiliation{Laboratoire de Physique des Solides - UMR 8502, CNRS,  Universit\'{e} Paris-Saclay, 91405 Orsay, France}

\begin{abstract}

Weakly flexible, directed polymers can adopt a potentially infinite variety of conformations, some of them  chiral.  We develop a second-virial theory for semi-rigid polymers embedded in a nematic environment based on the assumption that the chains possess no intrinsic molecular chirality but can take on weakly helical conformations of either handedness.  We demonstrate how these transient helical fluctuations can be exploited to predict the degree of chain stiffening induced by molecular crowding. The theoretical predictions are in good agreement with large-scale molecular dynamics simulations of bead-spring polymers with tunable backbone flexibility. Although both theory and simulations rule out spontaneous global chiral symmetry breaking imparted by  chain conformations alone, they do reveal that polymers can  develop pronounced enantiomeric shapes. Nematic fluids of semi-rigid polymers can be viewed as compensated cholesterics composed of transiently helical polymers with zero enantiomeric excess. By quantifying the degree of compensated chiral order developed over a broad range of concentrations and persistence lengths,  two distinct regimes of weak and strong enantiomerism can be identified. A crossover between the two occurs when the chain persistence length drops below roughly three times the contour length.
\end{abstract}

\date{\today}

\maketitle

\section{Introduction} 

Shape-persistent polymers capable of forming liquid crystalline  (LC) structures, characterized by long-range orientational order but with no full 3D crystallinity, are relevant for a wide range of soft matter materials. Examples are liquid-crystal devices \cite{uchida2022advanced,dkabrowski2004new}, nematic elastomers \cite{terentjev2025liquid},  emulsions \cite{concellon2023liquid}, droplets \cite{lopez2011drops} as well as living matter \cite{mitov2017cholesteric}  (for instance, 
cytoskeleton networks, neurofilaments within axons \cite{saw2018biological,safinya2013liquid}).

In crowded conditions, the polymer conformations are strongly correlated with those of neighboring chains. This is particularly true in a nematic liquid crystal where the polymer end-to-end vectors are collectively aligned along a common director \cite{de1993physics}. When the directional memory of chain segments decorrelate over a distance (called the persistence length) less or equal than the polymer contour length, the chains are referred to as semi-flexible. The effect of chain flexibility on, for instance, the liquid crystalline disorder-order transition has been the subject of a vast body of research usually based on the worm-like chain (WLC) as a benchmark model \cite{doi1988theory,marantan2018mechanics}. When bond angle fluctuations decay over distances much larger than the polymer length, the polymer can be effectively considered rigid and their phase behavior can be expected to be qualitatively similar to ``Onsager" rods with a length-to-width aspect ratio strongly exceeding unity \cite{onsager1949effects}. We refer the reader to a number of excellent articles reviewing the various theoretical approaches, and simulations dealing with such hard convex body models \cite{allen1993hard,mederos2014hard}.  

While both semi-flexible and rigid cases have been exhaustively explored, most recently in the context of geometric confinement \cite{chen2016theory,nikoubashman2021ordering}, important unanswered questions remain  as to what happens when polymers are {\em semi-rigid}, that is, when the persistence length amounts to many times the contour length but does not reach infinity such as for a perfectly rigid colloidal rod. In these situations, even though conformational fluctuations are much weaker than for the semi-flexible case they nevertheless contribute to the total entropy of the system and strongly impact their LC self-assembly, for instance in the case of reversible linear polymerization and living polymers \cite{van1994growth,wensink2019polymeric}.   

A common interpretation of semi-flexible polymer microstructure is based on chains being confined to cylindrical tubes made up of their neighboring particles. The characteristic length-scale is then  the so-called deflection length, which is much smaller than the persistence length, and is set by the typical distance along the chain over which the polymer is deflected back to the director due to presence of the confining tube formed by the neighboring chains. Recent simulations have pointed out the usefulness of this concept when these deflections are viewed as collective excitations modes \cite{egorov2016anomalous} beyond the original mean-field interpretation \cite{odijk1986theory}. 

For semi-rigid polymers, however, the deflection length usually exceeds the polymer size and thus no longer makes sense as a key length scale governing chain correlations in a nematic fluid. Instead, the polymers are taken as individual linear objects exercising only weak random conformational fluctuations.  In many studies to date on marginally flexible colloidal biopolymers, such as filamentous {\em fd} virus, the colloids are simply considered fully rigid and the role of conformational freedom is either neglected  altogether \cite{grelet2014hard} or absorbed into an effective softness featured in the effective pair interactions \cite{dennison2011phase,shundyak2006theory}. In this paper, we take a closer look at the importance of these shape fluctuations by combining a simple algebraic theory with large-scale computer simulation of a coarse-grained bead-spring model for purely repulsive polymers with tunable backbone flexibility.  The key assumption of our model is to regard the polymers as linear chains capable of adopting transient helical conformations that are readily parameterizable. Even though the mathematical helix is only a subset of the many possible configurations a chain can adopt, we find that the approach is quite successful in predicting, for instance, the response of the chain stiffness due to nematic crowding.  Our focus is on slender polymers with a large contour length over segment thickness ratio and which only interact with one other through steep repulsive interactions. In keeping with Onsager's original model \cite{onsager1949effects}, we assert that phase transitions occurring in such assemblies are  governed by entropic forces alone for which we invoke a second-virial theory that treats polymer-polymer correlations at the pair-level in terms of their excluded volume. This allows us to fully correlate the conformational freedom of each chain  with the degree of nematic alignment they experience in response to changes in the polymer concentration and free-polymer rigidity set by their persistence length.   Since all polymer conformations are constrained to be helical,  the theoretical model enables us to explore the possible development of enantiomeric conformational excess which would lead to spontaneous chiral symmetry breaking and the formation of cholesteric order without the presence of a molecular chiral source at the single polymer level \cite{fallwensink_prl2026}. Simultaneously, in our simulations we can  analyze the occurrence and probability of transiently chiral motifs of  conformationally {\em unconstrained} chains, test the various model predictions, and analyze the various enantiomeric regimes as a function of the chain flexibility.  

This rest of this paper is structured as follows. In Section II we begin by laying out our basic model for shape-transient helical polymers. Next, we develop a theoretical framework to describe nematic fluids of such objects through suitable extensions of Onsager's second-virial theory originally conceived for perfectly straight rigid rods.  The computer simulations are discussed in detail in Section IV.  To test the predictive power of the model we first analyze the degree of chain stretching in a nematic liquid crystal due to chain-chain interactions (Section V). Next, we discuss the chain conformations and their probability in an effort to measure the degree of enantiomerism developed by the chains for different bending rigidities and concentrations (Section VI). We also discuss various regimes of racemic and compensated cholesteric order that may occur and draw a brief analogy with molecular enantiomers. We finish the manuscript with a brief outlook and some concluding remarks (Section VII).

\section{Semi-rigid polymer as a shape-transient helix}

\begin{figure}
\includegraphics[width=0.8\linewidth]{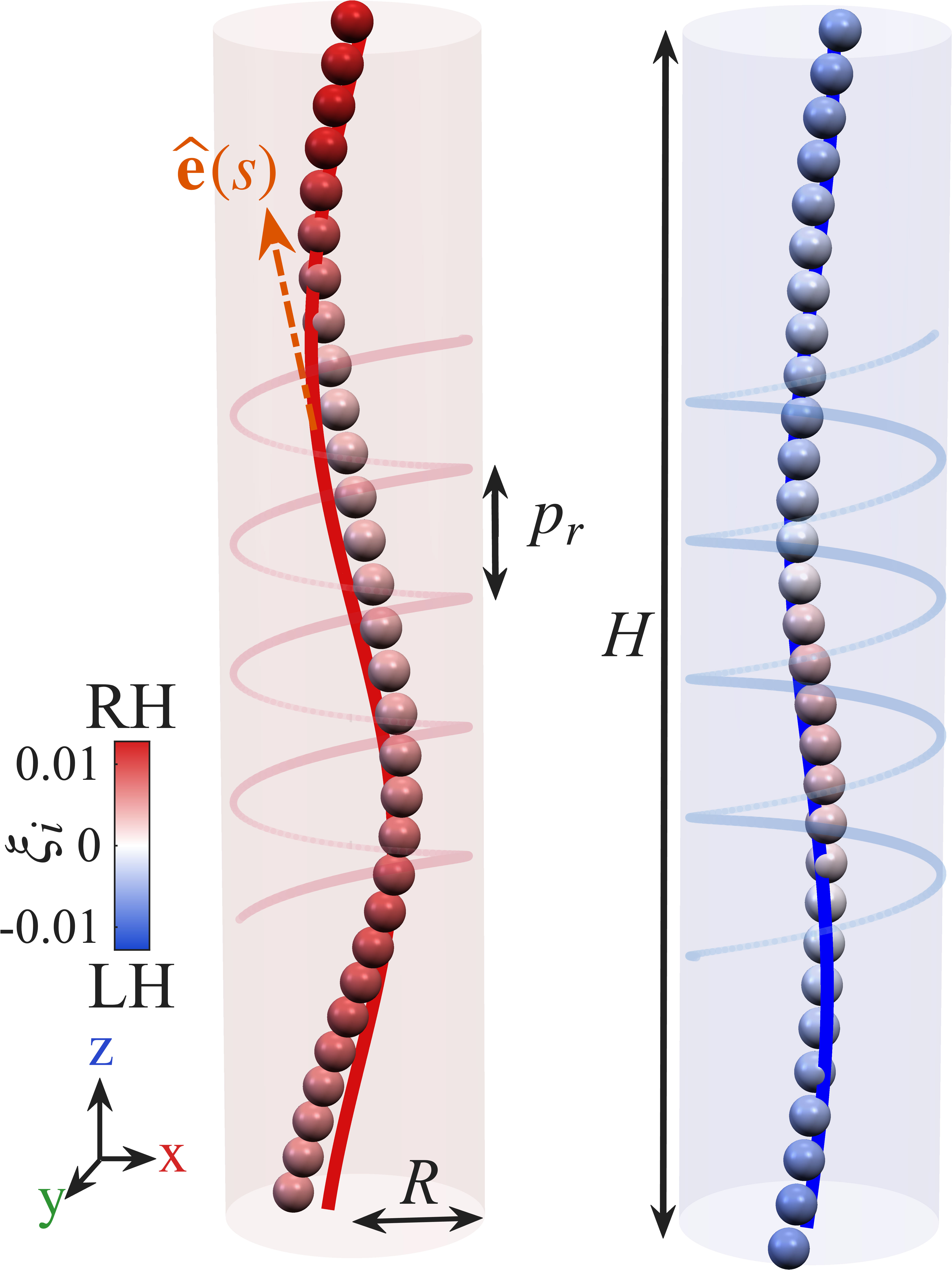}%
\caption{\label{schematic} Semi-rigid bead-spring chain represented as a best-fit mathematical helix [\eq{hepar}] with radius $R$, pitch $p_{r} = 2 \pi/q_{r}$, phase angle $\phi$ and Euclidian end-to-end distance $H$ (solid curve) for a predominantly right-handed chain and left-handed one. Beads are color-coded according to a local chirality parameter $\xi$ [\eq{intrachiral}] along the contour. }
\end{figure}







 Let us consider a weakly flexible polymer of length $L$ and diameter $D$ capable of adopting conformations of a strictly helical nature (see \fig{schematic}). The  position ${\bf p}(s)$ of the local chain segments is parameterized in a 3D Cartesian $(x,y,z)$ laboratory frame as
\begin{align}
{\bf p}(s) &=  {\bf p }_{c} + \left ( R\sin (q_{r} s + \phi) , R\cos (q_{r} s + \phi), s \right ),
\label{hepar}
\end{align}
with ${\bf p}_{c}$ the centre-of-mass position of the polymer and $s \in [-H,H]/2$ the position along the chain backbone and $H = L/\sqrt{1+ \chi^{2}}$ the Euclidian  end-to-end distance with  $\chi = q_{r} R$ a dimensionless parameter expressing the chirality of the helix in terms of a radius $R>0$ and  $q_{r} = 2 \pi/p_{r}$ in terms of the pitch distance $p_{r}$. Formally, $\chi = \kappa/ \tau$ in terms of the ratio between the helix curvature $\kappa  =q_{r}^{2}R/(1+ \chi^{2})$ and  torsion $\tau = q_{r}/(1 +\chi^{2})$. The phase angle $\phi$ is irrelevant for the single-helix properties described in this section but it will play a role when we discuss inter-helix correlations in Section III.    The handedness is determined by the sign of $q_{r}$ (and $\chi$) as per the usual convention. While $\chi =0$ is an obvious lower bound representing a perfectly straight chain, the upper limit for $\chi$ turns out to be a non-trivial problem for closed packed coils \cite{przybyl2001helical}. In practice, the space of available helix configurations, denoted by $(q_{r},R)$ is not unbounded but limited by self-avoidance of a tightly coiled close-packed chain. 

The conformational free energy per chain of length $L$ is given by the worm-like chain (WLC) energy \cite{doi1988theory} which for a perfect helix can be obtained in closed form
\begin{align}
F_{wlc} &= \frac{B}{2} \int_{-H/2}^{H/2} \frac{ds}{\sqrt{1+ \chi^{2}}} \left | \left | \frac{\partial {\bf \hat{e}}(s)}{\partial s} \right | \right | ^{2}  =  \frac{\frac{1}{2}B Lq_{r}^{2} \chi^{2}}{ (1 +  \chi^{2} )^{2}} \nonumber \\ 
& \approx \frac{1}{2}B Lq_{r}^{2} \chi^{2}    \hspace{0.5cm} (|\chi | \ll 1).
\label{fwlc}
\end{align}
The energy is proportional to the square-gradient of the tangential unit vector ${\bf \hat{e}}(s) = d {\bf p}/ds$ and vanishes for straight rods in two limits, namely $\chi \rightarrow 0 $ and $\chi \rightarrow \infty$. The first corresponds to the natural limit $R \rightarrow 0$ while the second limit is less physically relevant as it involves a vanishing pitch $q_{r} \rightarrow 0 $ with the radius $R$ simultaneously going to infinity. The combined effect is that $\chi \rightarrow \infty$ which leads back to a straight line.  For semi-rigid chains it suffices to consider contributions up to quadratic power in $\chi$. As the chain curls up the end-to-end distance $\mathcal{R}=H$ drops. For small $\chi$ we find
\begin{align}
 \mathcal{R}^{2}  &=  L^{2} \left ( 1 - \chi^{2} \right) + \mathcal{O}( \chi^{4} ).   
 \label{e2ehelix}
\end{align}
Introducing the persistence length $L_{P} = B/k_{B}T$ (with $k_{B}T$ the thermal energy in terms of Boltzmann's constant $k_{B}$ and temperature $T$) we quote the well-known Kratky-Porod result in the stiff chain limit ($l_{P} = L_{P}/L \gg L$) 
\begin{align}
\langle  \mathcal{R}^{2} \rangle_{\rm KP}  &= 2 L_{P}^{2} \left [ \frac{1}{l_{P}} - 1 + \exp \left ( -\frac{1}{l_{P}} \right ) \right ]  \approx L^{2} \left ( 1 - \frac{1}{3} \frac{1}{l_{P}} \right ). 
\label{krapo}
\end{align}
with the brackets $\langle \cdot \rangle $ denoting  a thermal average over all conformations. The corresponding probability function is given by the Boltzmann factor of the WLC energy
\beq 
f_{0}(q_{r}, R) = \mathcal{N}^{-1}\exp \left ( - \frac{1}{2}L_{P} Lq_{r}^{4} R^{2} \right ),
\label{fsingle} 
\eeq
Performing the conformation integral over the curvature $\chi$  greatly facilitates the analysis as the distribution becomes a Gaussian. The conformational average then becomes a single integral $\langle \cdot \rangle_{0} = \int d \chi (\cdot ) $. Further, to ease notation we implicitly normalize all variables in terms of the contour length. The normalization constant  becomes
\begin{align}
\mathcal{N} &= \int_{-\infty}^{\infty} d \chi \exp \left ( - \frac{1}{2}l_{P} q_{r}^{\ast 2} \chi^{2} \right ) = \left ( \frac{2 \pi}{l_{P} }\right )^{1/2}  \frac{1}{|q_{r}^{\ast}|},
\label{norm}
\end{align}
in terms of a typical pitch   $|q_{r}^{\ast}|$ that we can now specify  by mapping $\langle \mathcal{R}^{2} \rangle_{0}$ [\eq{e2ehelix}] using the mean-squared curvature $\langle \chi^{2} \rangle_{0} = 1/q_{r}^{2}l_{P}$  onto the Kratky-Porod result  \eq{krapo} in the limit semi-rigid limit $l_{P} \rightarrow \infty$. We thus find  $|q_{r}^{\ast}| = \sqrt{3}$ which corresponds to a chain pitch length $p_{r} = 2 \pi/\sqrt{3} \approx 3.6$ strongly exceeding the contour length. For simplicity, we shall define no upper bound on $\chi$ which means that rods can adopt random helical conformation with arbitrary short pitches in the limit $R \downarrow 0$. In practice, the upper limit for $q_{r}$ is typically set by the polymer  thickness $q_{r}^{\rm max} \sim D^{-1}$ but this extremum, in practice, is irrelevant in crowded nematic environments where the chains are strongly stretched. 

The mean-squared lateral extension of the chain is expected to scale  with the inverse persistence length and can be computed as follows.
First, in order to keep the chain centre fixed in space we   redefine the 2D lateral distance vector ${\bf p}_{ \perp}(s) = R \left ( -\sin q_{r} s, \cos q_{r} s \right )$ with respect to the helix mass centre via $\Delta {\bf p}_{\perp }(s) = {\bf p}_{ \perp }(s) - {\bf p}_{\perp c}  $ with ${\bf p}_{\perp c} = \tfrac{1}{H} \int_{-H/2}^{H/2} ds  {\bf p}_{ \perp }(s)$. In the limit of weak chain flexibility we find that radius of gyration perpendicular to the principal chain vector is given by 
\beq
R^{2}_{g\perp} =  \frac{1}{H} \int_{-H/2}^{H/2} ds || \Delta {\bf p}_{\perp} (s) ||^{2} \sim \frac{1}{12} \chi^{2}, 
\label{r20av}
\eeq
which vanishes at zero pitch $\chi \rightarrow 0$ as it should. In order to gauge the dispersion of the lateral chain extension with $q_{r}$ we compute the conformational fluctuation spectrum  which becomes
\beq
S(q_{r})  = \langle R_{g \perp }^{2} \rangle_{0R}  \sim \frac{1}{12 l_{P} q_{r}^{4}},
\label{specfree}
\eeq
suggesting that short-wavelength chain fluctuations are more strongly damped than long-ranged ones. This result is fully consistent with a general description of worm-like chain fluctuations discussed in the  Appendix. 



\section{Second-virial theory for transiently helical chains}

We proceed with describing a collection of chains forming a nematic fluid wherein the chain end-to-end vectors are strongly aligned but are non-parallel. Inspired by Onsager's considerations of strongly elongated rods solely interacting through volume exclusion we formulate the total free energy  in terms of a second-virial approximation based on pair interactions alone.  We define  $c = \rho v_{0}$ as the dimensionless chain concentration with $\rho = N/V$ the chain number density and $v_{0} = \pi L^{2} D/4$  the typical isotropic excluded volume of a perfectly stretched chain. The resulting Helmholtz free energy per chain reflects a balance of ideal, conformational and excluded-volume entropies \cite{onsager1949effects,vroege1992phase,mederos2014hard}
\begin{align}
& \frac{F[f]}{Nk_{B}T} = \int d \Omega d \chi  f(\Omega, \chi) [\ln ( c f(\Omega, \chi ) - 1] \nonumber \\ 
 + & \frac{3l_{P}}{2} \int d  \Omega d \chi f(\Omega, \chi)  \chi^{2} \nonumber \\ 
 + & \frac{c}{2} \int d \Omega_{1} d\chi_{1} f(\Omega_{1}, \chi_{1}) \int d \Omega_{2} d \chi_{2}    f(\Omega_{2}, \chi_{2}) \nonumber \\ 
 & \times v_{\rm ex} (\Omega_{1}, \Omega_{2}; \chi_{1}, \chi_{2}).  
\label{fv}
\end{align}
Here,  $\Omega$ denotes the solid angle of the helical object and the curvature $\chi =  q_{r}  R$  provides a single parameter measure for the conformational phase space [see \eq{norm}]. Of key interest  is the probability distribution $f(\Omega, \chi)$ which for a single chain  is set by the WLC energy \eq{fsingle} irrespective of the solid angle $\Omega$ of the helix.  For the many-body system under consideration, however, it is a priori unknown and  must be determined variationally as a minimum of the free energy for a given concentration $c$ and persistence length $l_{P} = L_{P}/L$. 

The above problem is a non-trivial one because  the excluded volume $v_{\rm ex} = V_{\rm ex}/v_{0}$  between two  chains not only depends on their individual dimensions but also on their mutual orientation and individual helical configurations.  For helices  whose main orientation directions are not necessarily parallel this quantity is non-trivial and an exact computation is impossible \cite{wensink2015chiral}.  Here, we shall approximate this quantity starting from the well-known expression for perfectly stretched rigid chains of length $L$ and diameter $D$ with principal orientations $\bhu_{1}$ and $\bhu_{2}$, namely $V_{\rm ex}^{(0)} = 2 L^{2} D |\bhu_{1} \times \bhu_{2} |$ ($D \ll L$). This expression can be generalized for arbitrarily curved chains as follows   
\beq
V_{\rm ex} (\Omega, \Omega'; \chi, \chi') \sim \frac{D}{2} \int_{-H}^{H} ds_{1} \int_{-H}^{H} ds_{2} \left | \frac{d {\bf p}_{1} }{d s_{1}} \times \frac{d {\bf p}_{2}  }{ds_{2}} \right |. 
\label{vexcontour}
\eeq
It is readily verified from the parameterizations in \eq{hepar}  that $V_{\rm ex} (\Omega, \Omega'; 0,0) = V_{\rm ex}^{(0)} $ for straight chains. Next, we adopt an orthonormal particle-based frame $\{ \bhu_{\alpha}, \bhw_{\alpha}, \bhv \} $ for each chain $\alpha =1,2$ with unit vectors $\bhv = (\bhu_{1} \times \bhu_{2}) / |\sin \gamma |$,   and  $\bhw_{\alpha} = \bhu_{\alpha} \times \bhv$ and $\gamma$ the enclosed angle between the end-to-end unit vectors $\bhu_{\alpha}$. Then the normalized tangent vector along the helix backbone of each helix reads 
\beq
N_{\alpha} \frac{d{\bf p_{\alpha}}}{ds_{\alpha}} =   \chi_{\alpha} \cos q_{r\alpha} s_{\alpha} \bhv - \chi_{\alpha}\sin q_{r\alpha} s_{\alpha} \bhw_{\alpha} + \bhu_{\alpha},
\label{tangu}
\eeq
with normalization $N_{\alpha} = \sqrt{1+ \chi_{\alpha}^{2}}$. Using the orthonormal  properties of the vectors along with the relations $\bhw_{1} \times \bhw_{2} = \bhv \sin \gamma $ and  $\bhw_{1} \times \bhu_{2} = \bhw_{2} \times \bhu_{1} =   \bhv \cos \gamma$ we can work out the cross product between the tangent unit vectors for two helices
\begin{align}
N_{1} \frac{d {\bf p}_{1} }{d s_{1}} \times N_{2} \frac{d {\bf p}_{2}  }{ds_{2}} &= (v_{1}w_{2} \bhu_{2} - w_{1} v_{2} \bhu_{1})  + (v_{2} \bhw_{1} - v_{1} \bhw_{2})  \nonumber \\  
&+ (w_1 - w_2 ) \cos \gamma \bhv + (1 + w_{1} w_{2} ) \sin \gamma \bhv, 
\end{align}
where $v_{\alpha}$ and $w_{\alpha}$ denote the coefficients in front of the unit vectors in \eq{tangu}.  For the computation of the excluded volume we need the norm of the above vector that we expand to quadratic order in the radius $R$ and perform a double contour integral following  \eq{vexcontour}. This enables us to obtain an analytical approximation valid for weakly helical departures from the straight chain limit and for strong nematic alignment  $| \sin \gamma | \approx | \gamma | \ll 1$
\begin{align}
& \frac{V_{\rm ex}}{2L^{2}D}  \sim | \gamma | \left ( 1 -(\chi_{1}^{2} + \chi_{2}^{2}) \right ) \nonumber \\
&  +\frac{1}{4 |\gamma |} \left [ \chi_{1}^{2} +\chi_{2}^{2} - 2\chi_{1}\chi_{2} j_{0}(q_{r1}L) j_{0}(q_{r2}L) \cos( \Delta \phi ) \right ],
\label{vex1}
\end{align}
with $\Delta \phi  = \phi_{2} - \phi_{1}$ the    phase difference between two helices and $j_{0}(x) = \sin x /x$ a spherical Bessel function. For weak pitches these functions may be set to unity without introducing great error.     We further ignore subtle correlations between the helix tips which are of the order of the internal helix volume $\mathcal{O}(LD^{2})$. These contributions are deemed unimportant in a second-virial approximation for slender helices. 
The first contribution in \eq{vex1}, proportional to $|\gamma|$, reflects a trivial reduction of volume exclusion due to a shortening of the end-to-end distance for any $\chi \neq 0$.   The second contribution is a non-trivial one and features a  divergence for strictly parallel configurations ($\gamma =0$) which is due to helices interlocking which  creates local overlap of the helix segments. In general, the regime of  near-perfect alignment is in itself a highly improbable extremum and requires $\langle \langle | \gamma | \rangle \rangle_{G}^{2} \gg \langle \chi^{2} \rangle$ for our approximation to be internally consistent. From here on,  $\langle \cdot \rangle_{G}$ denotes an orientational average of the end-to-end distance vector.  The orientational phase space is fully described by the polar $\theta$ and azimuthal  $\phi$ angles parameterizing the principal (pitch) axis of the helical chain and phase difference $\Delta \phi$ describing rotations of the non-uniaxial helical chains around their pitch axes.  

We also conclude from \eq{vex1}  that steric repulsion between helices of arbitrary shape intimately correlates with the phase angle difference $\Delta \phi$ and suggests helices are preferably polar to minimize steric hindrance ($\Delta \phi =0$). Drawing an analogy with magnetic spin models we conclude that a pair of homochiral chains with $\chi_{1} \chi_{2} >0$ will experience minimal steric repulsion when the chains are in a ``ferromagnetic" configuration $\Delta  \phi =0$. Similarly, an enantiomeric pair of helices ($\chi_{1} \chi_{2} <0$) will minimize their mutual repulsion in an anti-parallel pair configuration suggesting ``anti-ferromagnetic" ordering. The comparison can, of course, not be exact given that the chains operate in a fluid medium, in contrast to the static framework of a spin lattice.

Next, we factorize $f(\Omega, \chi) = f_{G}(\theta)f(\chi,\phi)$ in terms of a Gaussian distribution function of the polar angle $\theta$ of principal helix axis and the nematic director $\bhu \cdot {\bf \hat{n}} = \cos \theta$ assuming uniaxial order. The theory of Gaussian orientational fluctuations for strongly aligned hard-rod nematics has been detailed previously (see for instance Refs. \cite{odijk1985theory, wensink2019effect}). The decoupling enables us to simplify the previously stated free energy \eq{fv} into a reference part for straight chains and an ``excess"  contribution featuring the  probability function $f(\chi, \phi)$ for chain conformations
\begin{align}
& \frac{F[f]}{Nk_{B}T}  \approx  \ln c \alpha -1  + \frac{c}{2}  v_{\rm ex}^{(0)} + \frac{3l_{P}}{2} \int d \chi d \phi  f(\chi, \phi)  \chi^{2} \nonumber \\
& +  \int d \chi d \phi  f(\chi, \phi) [\ln (  f(\chi, \phi ) - 1]  \nonumber \\ 
& +  \frac{c}{2} \int  d \chi_{1} d \phi_{1} f(\chi_{1}, \phi_{1}) \int  d \chi_{2} d \phi_{2}    f(\chi_{2}, \phi_{2}) [ v_{\rm ex} - v_{\rm ex}^{(0)} ], 
\label{fv2}
\end{align}
with 
\beq
v_{\rm ex}^{(0)}  = \frac{8}{\pi} \langle \langle | \gamma | \rangle \rangle_{G}  \sim 8  /\sqrt{\pi \alpha},
\eeq
the excluded volume of straight  hard rods  ($\chi =0$) as per Onsager's original model in the Gaussian limit of strong alignment. The  parameter $\alpha \gg 1$ is a measure for the degree of nematic order under the proviso that the uniaxial nematic order of the fluid is retained. It follows from a minimum condition of the free energy as we will discuss later.    

\begin{figure}
\includegraphics[width=\columnwidth]{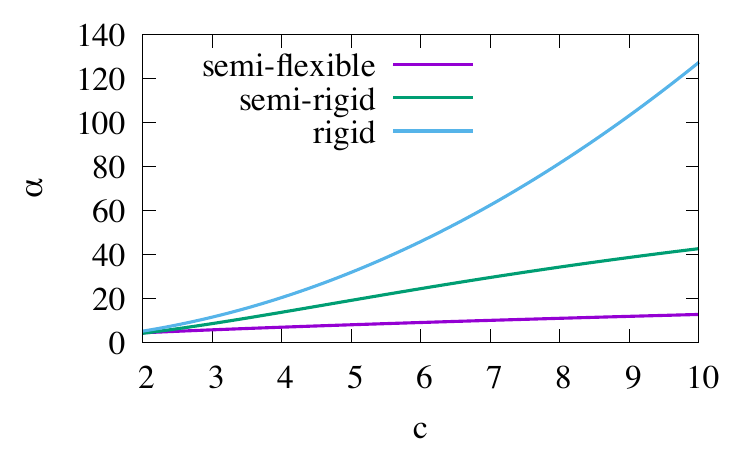}%
\caption{\label{alpha} Gaussian nematic order $\alpha$ expressed by $\alpha$ as a function of the chain concentration $c$ for semi-flexible and semi-rigid polymers (our model \eq{alfaintermediate}, $l_{P}=10$) and perfectly rigid rods ($l_{P} \rightarrow \infty$). 
 }
\end{figure}

Close inspection of \eq{vex1} reveals that the excluded volume is factorizable with the aid of the basic trignometric relation $\cos (\Delta \phi)= \cos \phi_{1} \cos \phi_{2} $ and ignoring the sine contribution. Further,  to ease notation we introduce  order parameters measuring (a)polar order of the phase angles for each conformation
\beq
\mathcal{S}_{n} = \int d \chi d \phi f(\chi, \phi) \chi^{n} \cos n \phi,  
\label{op}
\eeq
where $n=1$ measures polarity and $n=2$ nematic order. With these definitions the  above free energy can  be compactly expressed as (ignoring constants)
\begin{align}
& \frac{F}{Nk_{B}T}  \approx  \ln c \alpha   + \frac{c}{2}  v_{\rm ex}^{(0)} + \left (\frac{3l_{P}}{2}  + c \mathcal{A} \right ) \langle  \chi^{2} \rangle \nonumber \\ 
& +  \langle \ln   f(\chi, \phi)  \rangle  -\frac{c}{2} \mathcal{B}  \mathcal{S}_{1}^{2}.  
\label{fv3}
\end{align}
The coefficients $\mathcal{A}$ and $\mathcal{B}$ related to double averages over the end-to-end unit vectors of a chain pair
\begin{align}
{\mathcal A} &=  \frac{8}{\pi}  \left [ \frac{1}{4} \langle \langle \gamma^{-1} \rangle \rangle_{G}-  \langle \langle \gamma \rangle \rangle_{G}   \right ],  \nonumber \\ 
{\mathcal B} &= \frac{4}{\pi} \langle \langle \gamma^{-1} \rangle \rangle_{G}, 
\end{align}
so that $\mathcal{A} < \mathcal{B}$. Here we have made use of the standard Gaussian angular averages worked out by Odijk \cite{odijk1985theory}
\begin{align}
\langle \langle \gamma \rangle \rangle_{G} &\sim \sqrt{\pi/\alpha}, \nonumber \\
\langle \langle  \gamma^{-1} \rangle \rangle_{G}  &\sim \frac{1}{2} \sqrt{\pi \alpha}.
\end{align}
Next, a functional minimization of the free energy $\delta F / \delta f =0 $ yields an explicit expression for the conformational probability at finite polymer concentration $c$
\beq
 f(\chi, \phi) =\mathcal{N}^{-1} \exp \left [-\left ( \frac{3l_{P}}{2}  + c\mathcal{A}  \right ) \chi^{2}  + c \mathcal{B}  \mathcal{S}_{1} \chi   \cos \phi \right ], 
 \label{f0polar}
 \eeq
with $\mathcal{N}$ ensuring normalization of the distribution on the domains $\phi \in [0,2 \pi]$ and $ | \chi| \in [0, \infty \rangle$.  The polar distribution of the phase angles follows from the self-consistency condition
\beq
\mathcal{S}_{1}= \frac{ \int d \chi \chi \exp \left [-\left ( \frac{3l_{P}}{2}   + c\mathcal{A}  \right ) \chi^{2}   \right ] I_{1}(c \mathcal{B} \mathcal{S}_{1}\chi)} {\int d \chi \exp \left [-\left ( \frac{3l_{P}}{2}   + c\mathcal{A}  \right ) \chi^{2}   \right ] I_{0}(c \mathcal{B} \mathcal{S}_{1}\chi)},
 \label{s1sc}
\eeq
with $I_{n}(x)$ denoting modified Bessel functions. Expanding the right-hand side for $\mathcal{S}_{1}$ and performing the integration over $\chi$ we obtain the following solution
\beq
\mathcal{S}_{1} =
\begin{cases}
 0, & {\rm trivial} \hspace{0.1cm} {\rm solution}   \\
 \propto \pm \sqrt{c\mathcal{B} \langle \chi^{2} \rangle - 2 } , & {\rm if} \hspace{0.2cm} c\mathcal{B} \langle \chi^{2} \rangle >2.  
 \label{phaselock}
\end{cases}
\eeq
It turns out that $\mathcal{S}_{1} =0$ is the most relevant solution given that $c \mathcal{B} \langle \chi^{2} \rangle \approx c \mathcal{B} /3 l_{P} \ll 1$ for semi-rigid polymers. This can be intuitively understood from the notion that  departures from the straight uniaxial chain are too marginal to occasion a stable inhomogeneous distribution of phase angles driven by a balance between mixing entropy (favoring random phase angles) and chain volume exclusion that might be optimized under phase locking.  Even under conditions where phase locking is non-zero ($\mathcal{S}_{1} \neq 0$) there is no net global polarity transverse to the nematic director as the preferred phase angles of left and right-handed conformations and their associated polarity exactly cancel out.  
The relevant distribution thus becomes a simple Gaussian
\beq
f(\chi) =\mathcal{N}_{0}^{-1} \exp \left [-\left ( \frac{3l_{P}}{2}    + c\mathcal{A}  \right ) \chi^{2} \right ], 
 \label{f0even}
\eeq
with normalization $\mathcal{N}_{0} = \pi^{1/2} / (\tfrac{3}{2}l_{P} + c\mathcal{A})^{1/2}$. The nematic order parameter $\alpha$ can be determined from the  minimum condition of the free energy  \eq{fv3} that we recast in the following compact form
\begin{align}
\frac{F}{Nk_{B}T} &\sim \ln c \alpha  +\frac{4c}{\pi}\left (\frac{\pi}{\alpha} \right )^{1/2} +   \left ( \frac{3l_{P}}{2} +c\mathcal{A} (\alpha) \right ) \langle \chi^{2} \rangle_{0}
\nonumber \\
& + \langle \ln f_{0}(\chi) \rangle_{0}.  
\label{fvgauss}
\end{align}
The different entropy contributions are now easily identified; the first three  combine the ideal, orientational entropy and volume exclusion entropy of rigid rods. The last three feature conformational averages and respectively denote the effects of volume exclusion,  worm-like chain bending and conformational entropy introduced by marginal chain flexibility. In order to keep our analysis tractable we  approximated the conformational averages using the free chain  distribution $f_{0}$ \eq{fsingle}  so that the mean-squared curvature $ \langle \chi^{2} \rangle \approx \langle \chi^{2} \rangle_{0}  = 1/3 l_{P}$. By applying the minimum condition $d F/d \alpha = 0$ we obtain the following algebraic solution
\beq
\alpha_{SR} \sim 12 l_{P} - 8  + \frac{18 \pi}{c^{2}} l_{P}^{2} - \frac{6 \sqrt{4 \pi l_{P}^{2} (3 l_{P} -2) + 9 \pi^{2} l_{P}^{4}} }{c^{2}}.
\label{alfaintermediate}
\eeq
As expected, the nematic order for semi-rigid (SR) polymers exhibits a much weaker dependency on concentration than for perfectly rigid rods in which case it is quadratic
\beq
\alpha_{R} \sim 4c^{2}/\pi.
\eeq
The difference is due to the fact that flexible chains lose a considerable amount of configurational entropy upon aligning.  Since the degree of chain flexibility is only considered marginal in our model the concentration dependence is stronger than what is predicted for semi-flexible (SF)
rods ($l_{P} \approx 1$) \cite{vroege1992phase}
\beq
\alpha_{SF} \sim 4c^{2/3}/\pi^{1/3}.
\eeq
The results are graphically depicted in \fig{alpha} where we observe that the nematic order developed at a given concentration indeed resides between that of semi-flexible polymers and  perfectly rigid rods.

\section{Methods: Molecular dynamics simulation}
To test the theoretical predictions, we perform molecular dynamics (MD) simulations of semiflexible Kremer-Grest bead-spring chains using LAMMPS \cite{plimpton1995fast,thompson2022lammps}. The simulated system contains $N_{\mathrm{c}}=8001$ chains, each composed of $n=30$ spherical beads of diameter $\sigma$, corresponding to a contour length $L\simeq30\sigma$. Chain connectivity is imposed through finitely extensible nonlinear elastic (FENE) bonds, while excluded-volume interactions are represented by the purely repulsive Weeks--Chandler--Andersen (WCA) potential
\beq
U_{\mathrm{WCA}}(r)=4\epsilon\left[\left(\dfrac{\sigma}{r}\right)^{12}-\left(\dfrac{\sigma}{r}\right)^6+\dfrac{1}{4}\right],
\eeq
at $r\leq r_{\mathrm{c}}=2^{\frac{1}{6}}$ and $U_{\mathrm{WCA}} = 0$ at $r>r_{\mathrm{c}}$, where $r$ denotes the interparticle separation and $\epsilon$ sets the energy scale. Thus, the beads interact through short-ranged steric repulsion without any attractive contribution. The FENE interaction between consecutive beads is defined as 
\beq
U_{\mathrm{FENE}}(r)=-\frac{1}{2}k_{\mathrm{F}}R_0^2\ln\left[1-\left(\frac{r}{R_0}\right)^2\right ],
\eeq
where $k_{\mathrm{F}}=30k_BT/\sigma^2$ and $R_0=1.5\sigma$ denote the FENE spring constant and the maximum bond extension, respectively. Chain flexibility is controlled by a harmonic bending potential acting on every three consecutive beads of a rod molecule. This interaction is expressed as $U_{\mathrm{angle}}(\theta)=k_{\mathrm{angle}}(\theta-\theta_0)^2$, with the preferred angle $\theta_0=\pi$, where the rigidity of the chain is controlled by $k_{\mathrm{angle}}$. We investigate four bending rigidities, ranging from flexibility on the scale of the contour length to the semi-rigid limit: $k_{\mathrm{angle}}=8$, $50$, $100$, and $200~k_{\mathrm B}T/\mathrm{rad}^2$. A comparison with the continuum bending energy gives a bending modulus $B=2\sigma k_{\mathrm{angle}}$ and hence a bare persistence length $L_P=B/k_{\mathrm B}T=2\sigma k_{\mathrm{angle}}/k_{\mathrm B}T$. Using $L\simeq30\sigma$, the corresponding reduced persistence lengths are $l_{P} \simeq 0.5$, $3$, $7$, and $13$, respectively.  The three largest values represent semi-rigid rods with very weak backbone flexibility near the rigid rod limit. Chains with $l_{P} \simeq0.5$ are semi-flexible which falls outside the  scope of the present theoretical description for semi-rigid particles, but the case is kept as a useful reference system nevertheless. 

All simulations are performed in reduced LJ units, with $\sigma=1$ and $\epsilon=k_{\mathrm B}T$ defining the characteristic length and energy scales. The MD time is $\tau=(m\sigma^2/\epsilon)^{1/2}$, with timestep $\Delta t=0.005\tau$ and temperature  set at $T=1$. For each bending rigidity, the system is initialized by placing the $8001$ chains at random positions and random orientations in a sufficiently large dilute simulation box. The initially dilute configuration is first isotropically compressed under isobaric (NPT) conditions to a pressure $p=0.05 ~\epsilon/\sigma^3$. The resulting low-pressure system is subsequently simulated in the constant-volume (NVT) ensemble for $10^8$ MD steps, in order to relax the chain conformations and obtain an equilibrated isotropic reference state before the main compression protocol. To investigate the isotropic to nematic transition, we start from the equilibrated isotropic configurations and gradually compress the system in the NPT ensemble by continuously increasing the imposed pressure from $p_{\mathrm{start}}=0.05~ \epsilon/\sigma^3$ to $p_{\mathrm{stop}}=1.0~\epsilon/\sigma^3$ over $10^9$ MD steps. The barostat damping time is set to $p_{\mathrm{damp}}=10^{5}\Delta t$, ensuring a relatively slow response of the simulation box and suppressing rapid volume fluctuations during compression. The rate of compression is determined by the imposed pressure ramp and its duration, whereas $p_{\mathrm{damp}}$ controls the characteristic timescale over which the simulation box responds to changes in the target pressure. 

To characterize the equilibrium structures independently of the continuously compressed NPT trajectory, ten representative pressure state points are selected along the compression pathway. At each selected pressure $p_i$, an independent NPT simulation is performed at fixed pressure, {\it i.e.,} with $p_{\mathrm{start}}=p_{\mathrm{stop}}$. Each state point is subsequently simulated for an additional $10^8$ MD steps under constant NPT conditions, allowing the system to relax at the prescribed pressure before its equilibrium structural and orientational properties are evaluated. Statistical averages are obtained from $200$ configurations sampled from the stationary portion of each trajectory. The same preparation, compression, equilibration, and sampling protocol is applied independently to all four bending rigidities, enabling us to quantify how nematic crowding renormalizes the effective chain persistence length relative to its isotropic state.

The principal quantity we are interested in is the effective chain persistence length $l_{P}^{\rm eff}$. Let us first define the bond position ${\bf p}_{i}= ({\bf r}_{i+1} + {\bf r}_{i})/2$  (with ${\bf r}_{i}$ the position of bead $i$) and bond unit vector ${\bf \hat{e}}_{i} = ({\bf r}_{i+1} - {\bf r}_{i}) / ||  {\bf r}_{i+1} - {\bf r}_{i} ||  $. The persistence length denotes the length scale over which the tangent vector decorrelates and can be computed from the autocorrelation function
\beq
 \langle \bhe_{i} \cdot \bhe_{i+j }\rangle  = \exp \left ( - |j|\sigma/ L_{P}^{\rm eff} \right ), 
\eeq
where the brackets denote a thermal average. Alternatively, we can probe mean squared lateral extension of the chain. Let us first define the end-to-end-distance unit vector $\hat{\bf p} =\sum_{i} {\bf p}_{i} / || \sum_{i} {\bf p}_{i}|| $. Next, we project the bond positions in the molecular frame via $(\hat{\bf p} \otimes \hat{\bf p}) \cdot {\bf \hat{p}}_{i} = ( {\bf p}_{i}^{\perp}, p_{i}^{\parallel} )$. Then, the mean-squared lateral extension of the chain follows from
\beq
\langle R_{g \perp}^{2} \rangle  = \left \langle \frac{1}{2n^{2}} \sum_{i\neq j}^{n}  || {\bf p}_{i}^{\perp} - {\bf p}_{j}^{\perp} || ^{2} \right \rangle,
\label{r02}
\eeq
which is identical to the 2D radius of gyration of a directed polymer projected onto the plane perpendicular to the end-to-end distance vector.

Following Ref. \cite{fallwensink_prl2026} we can probe chiral chain conformations from a pseudo-scalar order parameter considering all tangent vector pairs and their distance unit vector $\Delta \hat{\bf p}_{ij}$ along each chain
\beq
\xi  = \frac{1}{{n}^{2}}  \sum_{i \neq j}^{n} ( {\bf \hat{e}}_{i} \cdot {\bf \hat{e}}_{j} ) [({\bf \hat{e}}_{i} \times {\bf \hat{e}}_{j}) \cdot \Delta \hat{\bf p}_{ij}]. 
\label{intrachiral}
\eeq
For a perfect helix in the weak curvature limit the order parameter can be approximated as $\xi \approx  \tfrac{1}{20} q^5 R^2 L^{3} $.  This quantity is trivially zero for a perfectly straight chain or for strictly achiral conformations such as a planar bent-core shape, but becomes non-zero when  chiral motives (e.g. local helical twist) emerge along the chain. An alternative measure suitable, for instance, for unbranched chiral polymers reads \cite{grant2026quantifying, abraham2024molecular}  
\beq
\xi  = \frac{1}{n} \sum_{i=3}^{n} ( {\bf \hat{e}}_{i-2} \times {\bf \hat{e}}_{i-1}) \cdot {\bf \hat{e}}_{i}, 
\eeq
which for a perfect helix with constant curvature gives a similar result, namely $\xi^{\prime} \approx q^{5} R^{2} ds^{3} $ where  $ds$  is an arbitrarily small length that can be identified with the bead size  $ds = \sigma$. Both measures essentially scale as $\xi \propto \kappa^{2} \tau$, that is, the square of the helix curvature $\kappa$ times the torsion $\tau$. 

Irrespective of the choice of $\xi$, the histogram $P(\xi)$ is expected to be an even function for a strictly racemic system indicating there is no prevalent handedness. However, the width and shape of the $P(\xi)$ may vary in a  non-trivial way which can help us distinguish various enantiomeric regimes as we will discuss in Section VI.

\begin{figure}
\includegraphics[width=\linewidth]{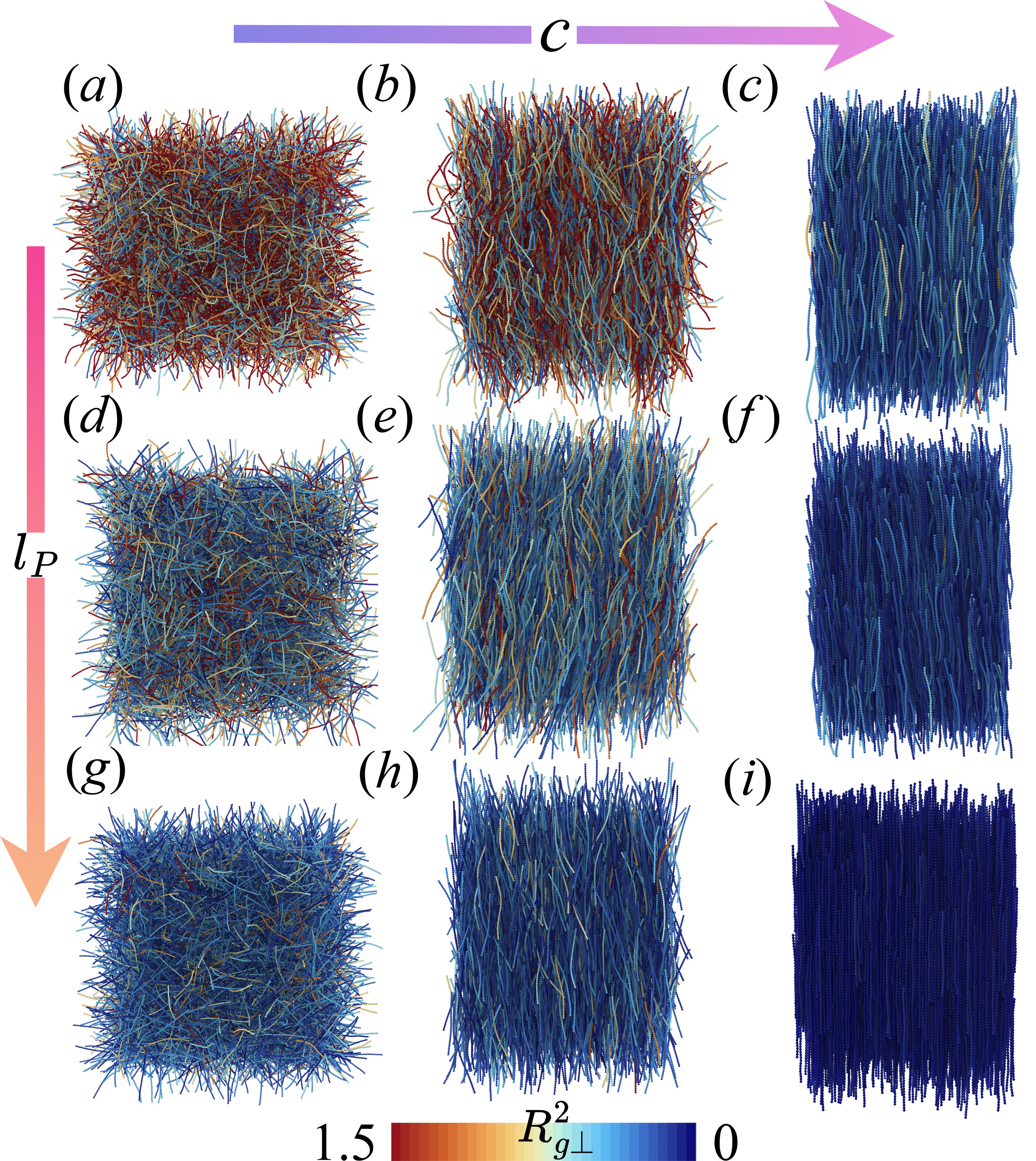}%
\caption{\label{simulation_snap} Simulation snapshots of semi-rigid polymers ($N\sim 8000$, each  consisting of $n=30$ beads) with persistence lengths $l_{P} = 3, 7, 13$ at concentrations $c = 2, 5, 14$. The corresponding  packing fractions are $\phi = 0.07, 0.17$ and $0.48$, respectively. The individual homopolymers are color-coded by the instantaneous lateral square gyration radius $ R_{g \perp}^{2}$ [\eq{r02}]. Note that the box volume varies across the snapshots.}
\end{figure}

  Typical simulation snapshots in \fig{simulation_snap} demonstrate standard isotropic and nematic fluid configurations with the lateral chain extensions color-coded. We note that at fixed concentration the nematic order increases with stiffness $l_{P}$ which is indicative of the isotropic-nematic transition moving to larger concentration as the chains become more flexible. This trend is well-known from previous studies on liquid crystalline semi-flexible polymers \cite{vroege1992phase, dijkstra1995simulation,egorov2016new}.

\section{Crowding induced chain stretching}

\begin{figure}
\includegraphics[width=0.9\columnwidth]{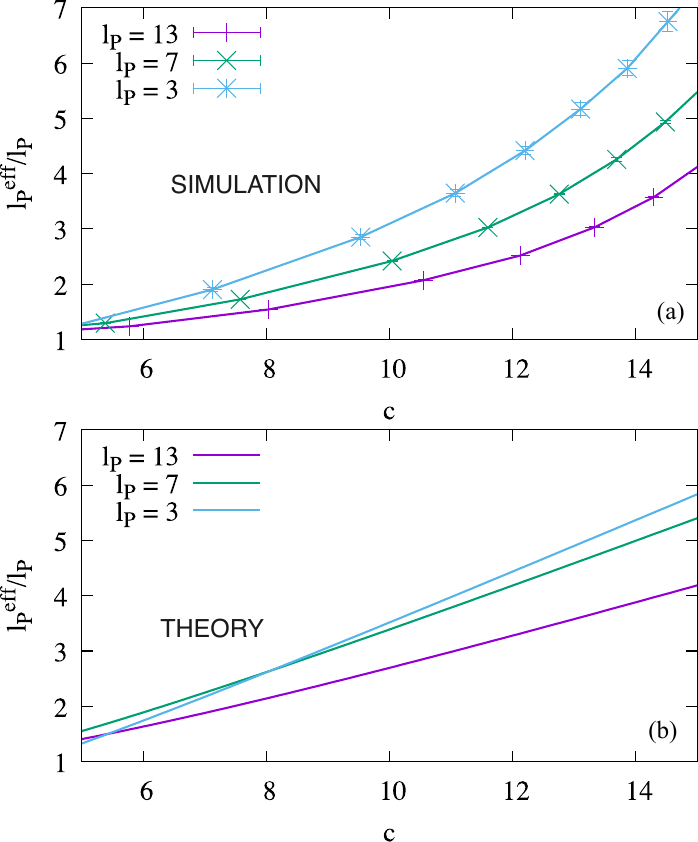}%
\caption{\label{perscomparison} Effective chain persistence length normalized by its free chain reference value as a function of concentration $c$ of the nematic fluid. Data are shown from simulations (a) and theory (b) for three different free chain persistence lengths $\l_{P} = L_{P}/L$ within the semi-rigid regime ($l_{P} \gg 1$).     }
\end{figure}

\begin{figure}
\includegraphics[width=\columnwidth]{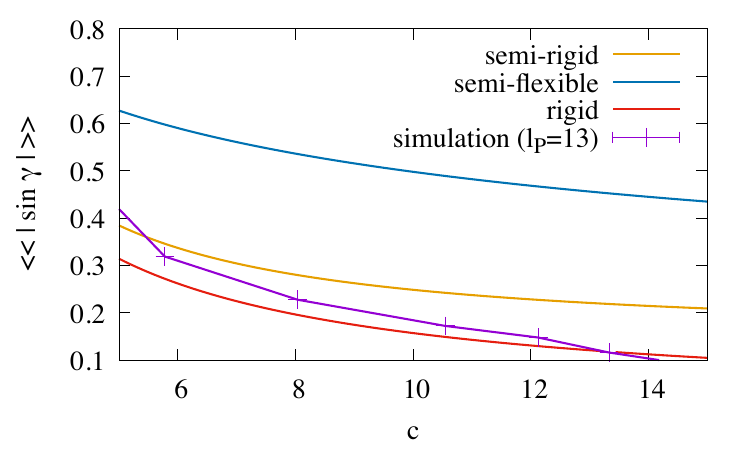}%
\caption{\label{gamma3} Evolution of $\langle \langle | \sin \gamma | \rangle \rangle$ as a measure of the average angle between chain end-to-end vectors as a function of the chain concentration $c$ for the case $l_{P} = 13$.  Qualitatively similar outcomes are obtained for the other persistence lengths.  }
\end{figure}

In order to get a first insight into the response of chain stretching at finite concentration  we consider the mean square end-to-end distance \eq{e2ehelix}.  By computing the mean-squared  curvature $ \langle \chi^{2} \rangle $ from the many-body chain conformational probability \eq{f0even} we find an analytical expression for the {\em effective} persistence length relative to its free chain reference value
\beq
\frac{l_{P}^{\rm eff}}{l_{P}} = 1 + \frac{2 \mathcal{A}c}{3 l_{P}}. 
\eeq
Note that this does not suggest a trivial linear increase with chain concentration as $\mathcal{A}$ depends implicitly on $c$ (and $l_{P}$) via the nematic order parameter $\alpha$ for which we take \eq{alfaintermediate}  for the semi-rigid case. 
Earlier on we formulated the criterion for undesirable near-parallel chain alignment in terms of the average angle between end-to-end chain vectors needing to be be greater than the typical later extension of the chains. Using the (Gaussian) averages obtained we conclude that this is generally the case for near-rigid chains at semi-dilute conditions ($ l_{P} \gg c$)
\beq
\frac{\langle \chi^{2} \rangle }{ \langle \langle | \gamma | \rangle \rangle_{G}^{2}} \sim \frac{4 c^{2}}{3 \pi^{2} } \ll 1.
\eeq
Combining \eq{r20av} and \eq{krapo} we can link the effective persistence length to the lateral gyration radius $\langle R_{g \perp}^{2} \rangle$ 
\beq
l_{P}^{\rm eff} \sim \frac{1}{36 \langle R_{g \perp}^{2} \rangle}.
\eeq
The latter is easily quantified from computer simulation previously discussed. A comparison between the theoretical and simulation predictions is shown in \fig{perscomparison}. While the  agreement is not fully quantitative, the overall trend namely a significant chain stiffening with concentration which gets more outspoken as $l_{P}$ lowers seems to be well captured by our  theory. The effective persistence length  increases almost about an order of magnitude from the low concentration region ($c \sim 5 $) just above the nematic-isotropic (N-I) transition up to deep into the nematic fluid.  As expected, the agreement between theory and simulation improves for stiffer chains.  In \fig{gamma3} the degree of nematic alignment of the chain end-to-end vectors is plotted in terms of the average angle between the chain end-to-end vectors. Clearly, semi-rigid chains align much more strongly than semi-flexible ones at a given concentration. 

\begin{figure}
\includegraphics[width=\linewidth]{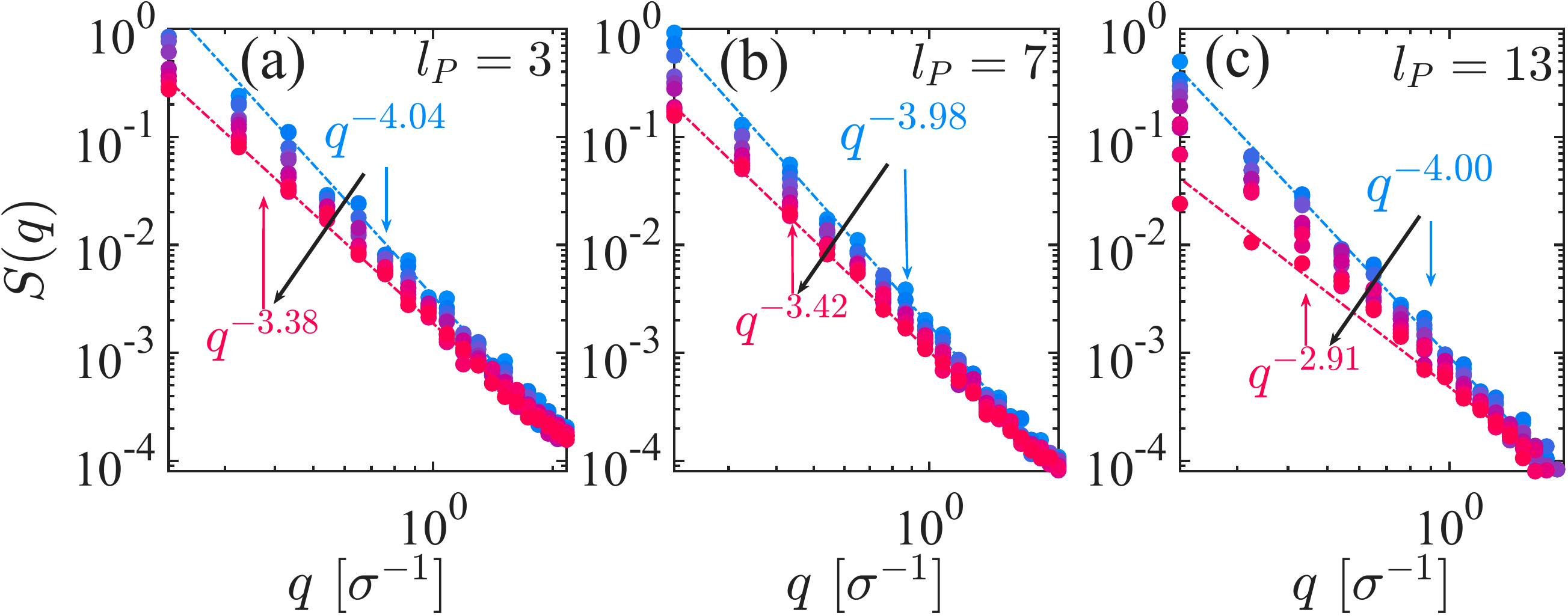}%
\caption{\label{fluctuations}  Dispersion of  transverse chain  fluctuations along the end-to-end vector for three persistence lengths $l_{P}=3 $ (a), $l_{P} = 7$ (b) and $l_{P} = 13$ (c)  as a function of the wavenumber $q$. Shown are results for the low-concentration isotropic system (in blue) featuring Porod scaling $S(q) \propto q^{-4}$ and the dense nematic regime (in red). }
\end{figure}

 The transverse fluctuation spectrum is equivalent to the mean square radius according to \eq{specfree}  and can be readily computed for correlated polymers from  \eq{r20av} using the  Gaussian distribution \eq{f0even} and integrating over the radius $R$. This leads to
\beq
S(q_{r}) \propto \frac{q_{r}^{2}}{\left (\tfrac{1}{2} l_{P} q_{r}^{4} + \mathcal{A} c q_{r}^{2} \right )^{3/2} }.
\label{speccrowded}
\eeq
The simulation data are shown in \fig{fluctuations}. Upon crowding the free chain scaling [$S(q) \propto q^{-4}$ for $c \rightarrow 0$] becomes less steep in accordance with the above prediction. Physically, it means that as lateral chain fluctuations get suppressed by crowding they also get more dispersed allowing for (helical) shapes with larger wavenumbers to become relatively more prominent.

\section{Conformational enantiomerism of semi-rigid polymers}

In view of the lack of a molecular chiral source in our chain model the conformational probability \eq{f0even} is an even function peaked at $\chi = 0$ suggesting that helical conformations of the L or R-handed type should be manifest at equal probability and that the system is a racemate. However, one could envisage a scenario whereby chains collectively pick one handedness such as to avoid unfavorable volume exclusion between  polymer of unlike symmetry. We have seen that such a spontaneous chiral symmetry breaking can not occur within the weak curvature regime considered in this work since  unfavorable volume exclusions are ``neutralized"  by a random distribution of the phase angles which is the entropically most favorable state. A second mechanism at play is that an enantiomeric chain pair can mitigate their unfavorable volume exclusion by stretching, thereby reducing their chirality $\chi$. This, in turn, costs conformational entropy.  The dominant state is therefore a racemic nematic phase composed of chains performing weakly helical conformational excursions of either helical sense.

From a  theoretical point of view, however, a different kind of racemate could be contemplated from a local free energy minimum as the result of a subtle trade-off between volume exclusion changes, namely, a reduction of the Euclidian length as the chains depart from their stretched ground state and a simultaneous increase of the effective thickness as $|\chi|$ grows. The conformational probability,  established  in \eq{f0even}, can be expressed as a Boltzmann distribution
\beq
f(\chi) \propto e^{- \mathcal{U}(\chi)/k_{B}T },
\eeq
in terms of  an effective potential of mean force $\mathcal{U}(\chi)$ as a function of the curvature $\chi$
\beq
\frac{ \mathcal{U}(\chi) }{k_{B}T}  \sim   M \chi^{2}   - 3 l_{P} \chi^{4} + \frac{9}{2} l_{P} \chi^{6}, 
\label{ueff}
\eeq
with $M = (3l_{P}/2) + c\mathcal{A}$. The quartic and sixth-order contributions emerge from the expansion of the worm-like chain energy \eq{fwlc}
\beq
\frac{F_{wlc}}{Nk_{B}T} \approx \frac{3l_{P}}{2} \left ( \chi^{2} - 2 \chi^{4} + 3 \chi^{6} +  \cdots \right ). 
\eeq
The term of $\mathcal{O}(\chi^{6})$ is necessary to keep the chain from collapsing into a highly-coiled object. These collapsed states are naturally prevented by the segment self-avoidance of a finite thickness helix as well through large deformation corrections to the worm-like chain energy that we do not consider here. 
In principle, the higher order curvature corrections to the WLC energy ought to be accompanied with their counterparts from the chain excluded volume [\eq{vexcontour}]. Although these could be systematically derived from the previous analysis they are not included here for the sake of argument.

\begin{figure}
\includegraphics[width=0.9\columnwidth]{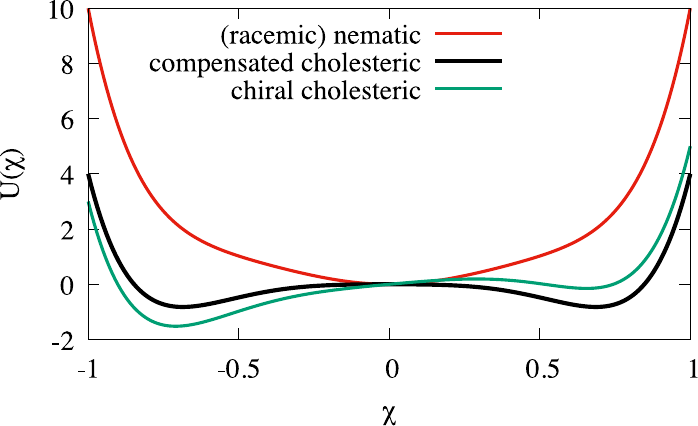}%
\caption{\label{mhat} Potential of mean force $\mathcal{U}(\chi)$ [\eq{ueff}] of a semi-rigid chain with $l_{P}=3$ under nematic crowding. Three possible scenarios are possible: a nematic ($c \mathcal{A}=1$), a compensated cholesteric ($c \mathcal{A}=-5$) where finite $|\chi|$ is favored but no specific handedness, and a conventional cholesteric characterized by chiral symmetry breaking ($ c\mathcal{A} = -5, \varepsilon=0.2)$  in which chains with negative symmetry ($\chi <0$) are biased over those with $\chi >0$ (or vice versa). The latter scenario follows from adding a linear term $ \propto \varepsilon \chi$ to \eq{ueff} [see also \eq{fvchiral}]. In our system, strictly,  $\varepsilon=0$ because of lack of a molecular chiral source.    }
\end{figure}

The potential $\mathcal{U}(\chi)$ may exhibit a local minimum which is determined from the extremum condition $d \mathcal{U}(\chi)/d \chi =0$. 
The potential then attains a Mexican hat shape with optimal curvature  $\chi_{m}$ featuring as a local potential minimum.  This is illustrated by the black curve in \fig{mhat}.  The minimum is located at
\beq
\chi_{m}^{2} = \frac{2}{9} \left ( 1 + \sqrt{1 - \frac{3M}{2l_{P}}} \right ).
\eeq
Physically relevant solutions are possible only if the coefficient $\mathcal{A}$ is negative,
$c \mathcal{A} < -\frac{5}{6} l_{P}$.
This case leads to strong enantiomerism as it is characterized by helical chains with opposite handedness in which the polymers preferentially adopt a finite curvature $\chi_{m}$ as stipulated by the local minimum. Consequently,  chains trapped in a potential minimum need a finite energy to cross over into a configuration of opposite handedness as they can only do so by unwinding ($\chi \rightarrow 0$) and recoiling in the other helical sense. In the weakly enantiomeric regime, such energy barrier is absent.

The crossover between the two regimes only makes sense in context of the isotropic-nematic phase diagram that we need to determine first. Relevant expressions for the pressure and chemical potential are obtained from the standard thermodynamic derivatives of the nematic free energy \eq{fv3}. Its isotropic counterpart is given by 
\begin{align}
\frac{F_{\rm iso}}{Nk_{B}T} &\sim \ln c  +c + \left (\frac{3l_{P}}{2}  + c\mathcal{A}_{\rm iso} \right ) \langle \chi^{2} \rangle_{0} + \langle \ln f_{0}(\chi) \rangle_{0}, 
\label{fv4}
\end{align}
with $f_{0}$ given by the free-chain conformational probability \eq{fsingle}   
\beq
{\mathcal A}_{\rm iso} =  \frac{8}{\pi}  \left [ \frac{1}{4} \langle \langle \sin^{-1} \gamma \rangle \rangle_{G}-  \langle \langle \sin \gamma \rangle \rangle_{G}   \right ] = -1,
\eeq
using the isotropic averages $ \langle \langle \sin^{-1} \gamma \rangle \rangle_{G} = \pi/2$ and $\langle \langle \sin \gamma \rangle \rangle_{G}   = \pi/4$. For simplicity we assume  free-chain conformational averages for both isotropic and nematic phases meaning that the worm-like chain  energy and conformation entropy is assumed to be the same in both phases. In order to connect to the semi-flexible case we use results from the Gaussian approximation $c_{I} = 7.77/l_{P}$ and $c_{N}=9.71/l_{P}$ ($l_{P} <1 $) \cite{vroege1992phase}. This approach can, at best,  give a qualitative picture of the phase boundaries  because  a Gaussian angular distribution for the polymer end-to-end vectors is known to overestimate the isotropic-nematic transition compared to more accurate distributions \cite{vroege1992phase}.

\begin{figure}
\includegraphics[width=\columnwidth]{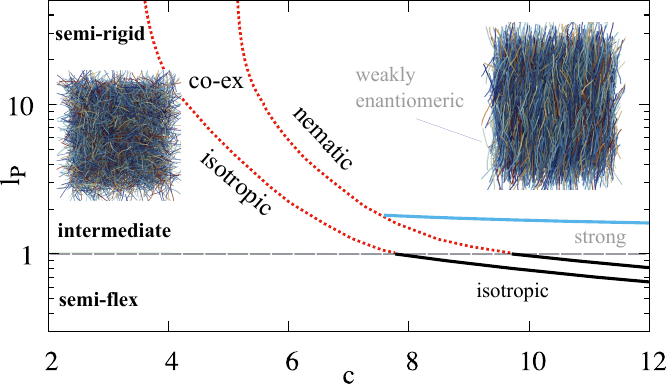}%
\caption{\label{pd}  Tentative phase diagram demonstrating the isotropic-nematic phase regions and enantiomeric regimes, plotted in terms of the chain concentration $c$ versus the free chain persistence length $l_{P}$. The regime between semi-flexible and semi-rigid features a crossover (in blue) from weak to strong enantiomerism.  The transition towards phase locking ($\mathcal{S}_{1} \neq 0$) [\eq{phaselock}] happens very close  to the crossover.  }
\end{figure}

Based on the above we are able to construct a tentative phase diagram  summarizing the main features [\fig{pd}].  The isotropic-nematic transition dramatically shifts to higher concentrations as the flexibility increases and is dominated by a conventional nematic order. Within the intermediate regime  when the chain persistence length is only a few times the contour length, a crossover to strong enantiomerism appears. A unimodal $P(\chi)$ reflects conventional nematic order of stiff polymers where the chains exercise weakly helical departures around a perfectly stretched reference state. When chain conformations become trapped in local minima with finite helical curvature $\chi$ the histogram should develop additional peaks or shoulders at finite $|\chi|$.

\begin{figure}
\includegraphics[width=\linewidth]{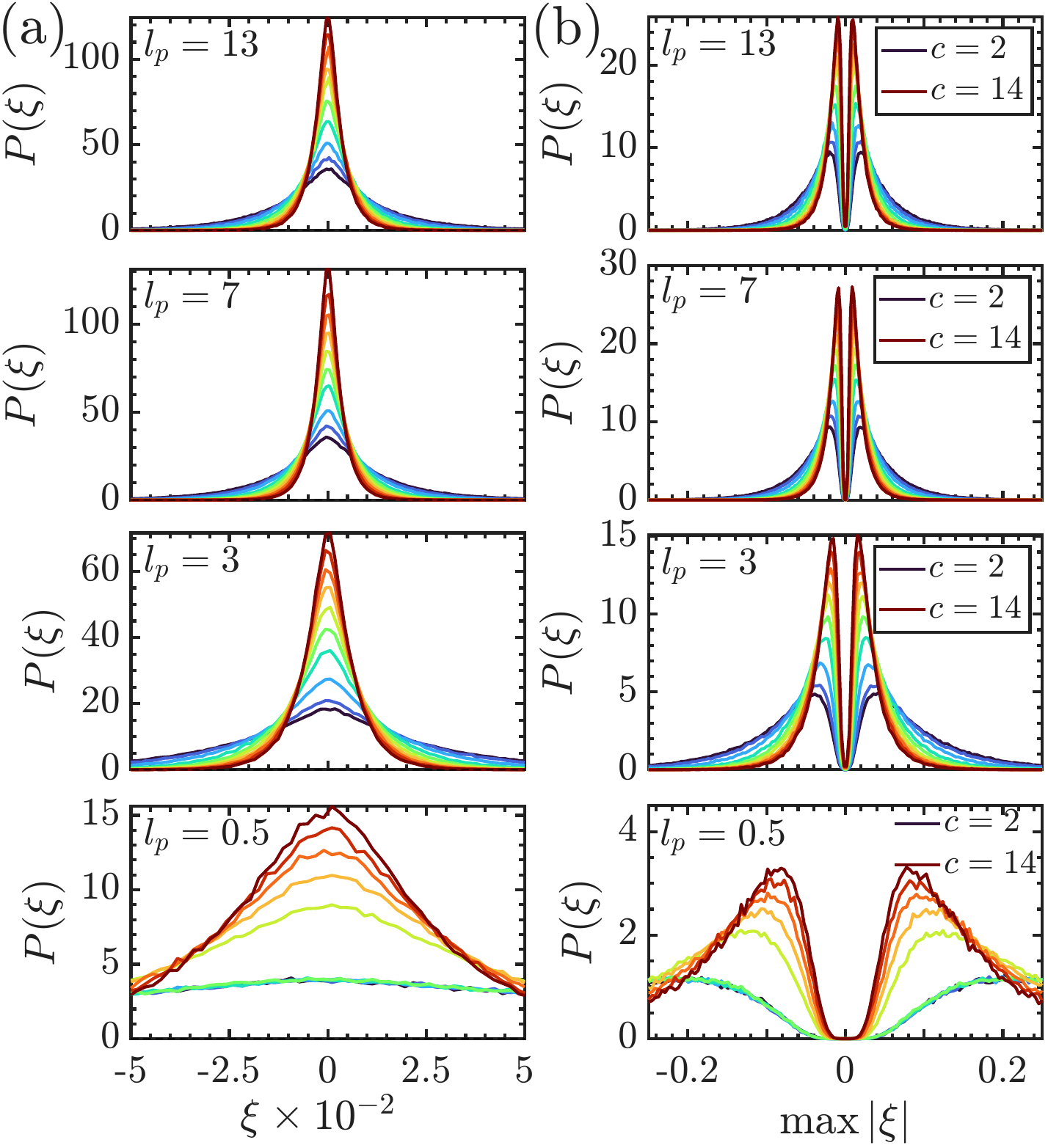}%
\caption{\label{histo}  Probability distribution  of the chiral order parameter $\xi$ [\eq{intrachiral}] for polymers at various persistence lengths $l_{P}$ and concentrations $c$ (color-coded) both in the isotropic and nematic phases. The panels on the right display the probability of the maximum value of $\xi$ developed following the contour $n \in [1,30]$ of each chain.  }
\end{figure}

\begin{figure}
\includegraphics[width=\columnwidth]{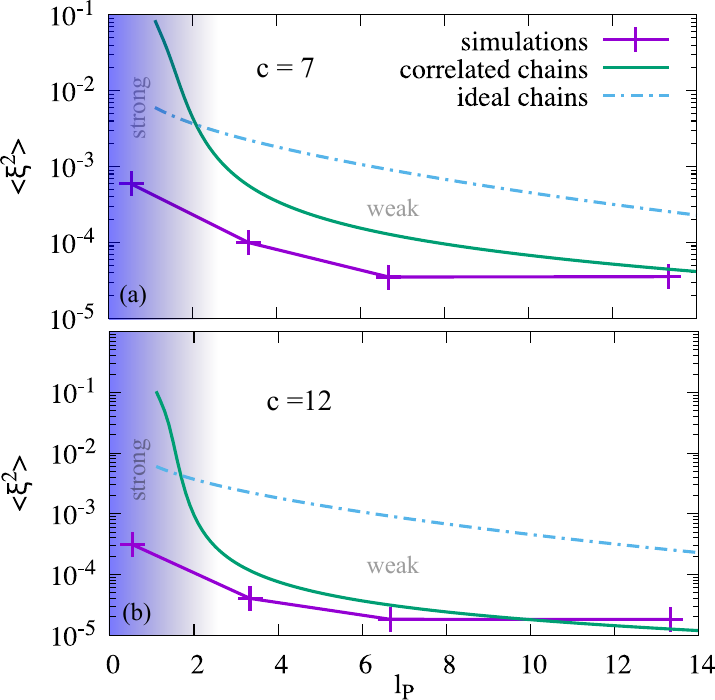}%
\caption{\label{variance}  Variance of the chiral order parameter $\xi$ [\eq{intrachiral}] as a function of the persistence length $l_{P}$  for a low concentration nematic phase with $c=7$ (a)  and a high concentrated one with $c=12$ (b). The shaded zone indicates a crossover from weak to strong shape enantiomerism.  }
\end{figure}

The histograms obtained from simulations are given in \fig{histo}. Here, chirality is expressed by the order parameter $\xi$ [\eq{intrachiral}] instead of the helix curvature $\chi$. Although the $\xi$  is too course a measure to reveal any shoulders appearing in $P(\xi)$, as predicted by the theory, the distributions widen as the flexibility increases allowing the chains to explore a wider range of chiral conformations. In line with the theoretical prediction, the variance exhibits a marked widening below a critical $l_{P} \approx 3$ [\fig{variance}]. This points to a crossover from weak to strong shape-driven enantiomerism. The notion that a nematic liquid crystal of weakly flexible chains is in fact composed of enantiomeric shapes is also reflected from the histograms of the maximum chiral order developed by each chain [right panels in \fig{histo}]. There is a marked optimum in the chiral strength at either handedness. The distance between the peaks  widens considerably as the persistence length approaches the semi-flexible range ($l_{P} \rightarrow 1$) pointing to a growing level of compensated chirality at decreasing chain rigidity.

\section{Discussion and conclusions}

Slightly flexible or semi-rigid polymers are never straight but tend to adopt seemingly helical conformations even though the curvature may vary along the chain contour.  By constraining the complex configurational phase space of a semi-rigid polymer to a simple, perfect helix with constant curvature but arbitrary pitch and radius we are able to develop an algebraic theory for interacting polymeric enantiomers forming a nematic fluid.  The approach enables us to address the role of conformation freedom for a class of nematogenic polymers residing between fully rigid rods and semi-flexible chains.  Crowding effects naturally occur in a nematic system of aligned polymers and the global alignment is demonstrated to lead to a stiffening of the polymer chains in terms of an enhanced effective persistence length. This effect is tested against computer simulations of a bead-spring polymer model and we find that the theory gives the correct rate of chain stiffening with concentration. 

Even though the polymer configurations are constrained to be helical, there is no spontaneous chiral symmetry breaking (CSB) which would imply the development of non-zero enantiomeric excess and the formation of cholesteric order as recently reported for the more specific case of polydisperse chains near the NI transition \cite{fallwensink_prl2026}.  The argument ruling out CSB for transiently helical polymers, apart from a trivial lack of molecular chirality, is that the volume exclusion between chains of unlike symmetry versus that of like chains is insufficient to drive such a transition. Spontaneous deracemization will only occur if like-handed conformations are sufficiently favorable to off-set the loss in mixing entropy associated with a transition from a racemic to a homochiral state. In  polymeric systems, such an effect could be brought about, for instance,  by the addition of chiral dopants that trigger chiral amplification \cite{green1999macromolecular, palmans2007amplification, deutsch2026chirality}.  Alternatively, the presence of non-directional intrachain interactions \cite{} such as those generated by the addition of small depletants \cite{snir2005entropically} or by varying solvent conditions \cite{biswas2021rigidity} might give rise to a double-well conformational energy landscape similar to the black curve  in \fig{mhat}.   

Our simulations reveal that the maximum chiral strength each chain and its enantiomeric counterpart is able to impart exhibits an optimum probability that grows with the backbone flexibility.  The nematic phase of semi-rigid polymers could be interpreted as a compensated cholesteric in terms of racemic mixtures of shape-driven enantiomers with no net macroscopic chirality. The compensation effect grows significantly stronger as the polymers become more flexible yet remain stiff enough to guarantee stable nematic order.

To the best of our knowledge compensated cholesterics have not been identified before in relation to lyotropic liquid crystals. The term is commonly used in the context of ``manufactured" mixtures of chiral molecular nematogens of different handedness and/or chiral strengths where the composition or temperature of the mixture is tuned such as to obtain a cholesteric with infinite pitch \cite{sackmann1968structure, kozawaguchi1978helical,katsonis2012controlling}. At the compensation point  the system exhibits no net chirality as measured through, for instance, optical rotation. In molecular systems, a slight variation of the temperature away from the compensation point leads to non-zero enatiomeric excess and the development of macroscopic chirality (cholesteric order). In the current system such symmetry breaking  could, for instance,  be achieved by adding a small amounts of intrinsically chiral polymers as dopants. Unlike most chiral molecular nematogens, the effective shape and hence the chiral strength of the polymer is not fixed but fluctuates about an average value dictated by the polymer shape that is controlled by the polymer concentration which also governs the degree of alignment. The enantiomeric nature of the nematic fluid remains relatively weak for semi-rigid polymers but becomes much stronger when the flexibility approaches the semi-flexible regime. A typical crossover value of $l_{P} =3$ was found.

As an outlook we want to briefly touch upon the situation when chains do have a chiral molecular source of some kind so that helical conformational motifs of one handedness are energetically preferable over the other \cite{yamakawa1997helical,lubensky1996chiral,pieraccini2010chirality}.  Within the current model this scenario could be addressed by including a linear term to the free energy that is proportional to some molecular chiral amplitude $\varepsilon$. Then, for a cholesteric liquid crystal the free energy per unit volume  reads \cite{de1993physics}
\begin{align}
\frac{ F_{\rm chiral}}{V} =  \frac{F}{V}   + \rho \varepsilon \langle \chi \rangle  +  \tfrac{1}{2} K_{2} ( \bn \cdot \nabla \times \bn + q_{0})^{2},  
\label{fvchiral} 
\end{align}
where $F$ is the Helmholtz free energy for the non-chiral reference system ($\varepsilon=0$) given by \eq{fv3}. The sign of $\varepsilon$ determines the chiral bias (L or R) and $\langle \chi \rangle$ expressing the enantiomeric excess of the polymers. The quantities $K_{2}$ and $q_{0}$ respectively refer to the twist elastic modulus and chiral pitch describing the helical twist of the local nematic director $\bn$. Both can be computed within the second-virial framework developed here by extending the theory to accommodate for nematic systems with weak spatial modulations of the director field \cite{straley1976theory}.  The results could appeal to a wide range of non-rigid biopolymers composed of intrinsically chiral molecular units including nanocrystalline chitin \cite{revol1993vitro}, cellulose \cite{lagerwall2014cellulose, tran2020understanding},  amyloid fibrils \cite{nystrom2018liquid,jin2024structural} or filamentous virus  \cite{grelet2003origin,grelet2024elucidating}. The case of chiral semi-rigid polymers, however,  necessitates a  more wide-ranging theoretical analysis that explicitly addresses  $K_{2}$ and $q_{0}$ and their response to changes in chain alignment and  rigidity. The key challenge would be to make a systematic exploration of the relation between microscopic chirality prescribed at the chain segment level set by $\varepsilon$ and the mesoscopic cholesteric pitch $q_{0}$ for a range of chain flexibilities, most notably in the semi-rigid regime ($l_{P} \gg 1$) which has received little attention thus far.  This will be the subject of a forthcoming study.

\section*{Acknowledgement}

This work was supported by the European Innovation Council (EIC) through the Pathfinder Open grant “INTEGRATE” (no. 101046333). The authors acknowledge HPC resources from GENCI-IDRIS (Grant 2024-[A0170913823] and 2025-[A0190916897]). 

\section*{Appendix: General description of worm-like chain fluctuations}

 The local unit tangent vector of a worm-like chain of length $L$ can be written as \cite{tortora2020chiral}
\beq
\bhe(s) = e_{\parallel} (s) \bhu + {\bf e}_{\perp} (s),  
\eeq
where $s \in [0,L]$ is the curvilinear abscissa, $\bhu$ the end-to-end vecto
r and $e_{\parallel}^{2}(s) + || {\bf e}_{\perp} (s)  ||^{2} = 1  $. For large $L_{P} \gg 1$ the transverse fluctuations are very small ($||  {\bf e}_{\perp} (s)  ||^{2} \ll 1 ||$) so that $e_{\parallel}(s) \approx 1$. The local chain segments can then be parameterized as
\beq
\br(s) \approx  \br(0) - \br_{\perp}(0) +   s \bhu  + \br_{\perp}(s). 
\eeq
Assuming periodic end conditions $\br_{\perp}(-L/2) = \br_{\perp}(L/2)$  we express the transverse fluctuations as a Fourier series 
\beq
\br_{\perp}(s) = \sum_{q} \hat{\br}_{\perp}(q) e^{i qs },  
\eeq
with discrete wavenumbers denoted by $q$. The coefficients are vectors parameterized as follows
\begin{align}
\hat{\br}_{\perp}(q) = \frac{1}{L}\int_{-L/2}^{L/2} ds \br_{\perp}(s) e^{-i q s }  = \hat{p}_{\perp 1}  \bhv + \hat{p}_{\perp 2}   \bhw,
\end{align}
in terms of the orthonormal molecular frame $( \bhu , \bhv, \bhw )$ and complex coefficients $\hat{p}_{\perp i} (q )$ ($i=1,2$). The free energy per chain in units of thermal energy can then be expressed as a sum over these coefficients
\begin{align}
 \frac{F_{wlc}}{k_{B}T} \approx \frac{L_{P}}{2} \int_{-L/2}^{L/2} ds \left | \left|  \frac{d^{2} \br_{\perp}(s)}{ds^{2}} \right | \right |^{2} = \frac{L_{P}L}{2} \sum_{q} q^{4}  || \hat{\br}_{\perp}(q) ||^{2}, 
\end{align}
with $ || \hat{\br}_{\perp}(q) ||^{2}  =  \hat{p}_{\perp 1}  \hat{p}_{\perp 1}^{\ast}   + \hat{p}_{\perp 2}  \hat{p}_{\perp 2}^{\ast}  $ where the asterisk denotes a complex conjugate.  From the equipartition theorem stating that all modes contribute an equal amount of $k_{B}T/2$ we find that the root-mean-squared transverse fluctuations for a free helical chain exhibits the following (Porod) scaling  
\beq
 S (q) = \langle || \hat{\br}_{\perp}(q) ||^{2} \rangle  \sim \frac{2}{L L_{P} q^{4}}.
\label{swlc}
\eeq
For a perfect helix considered in the model we  have the following parameterization
\beq
\br_{\perp}(s) = R [\cos (q_{r} s + \phi ) \bhv +   \sin (q_{r} s + \phi ) \bhw],
\label{helix}
\eeq
with $\phi$ an arbitrary phase angle. For simplicity we let the helices be infinitely long and obey periodic end conditions. Switching to Fourier space the modulus reads $ || \hat{\br}_{\perp}(q)||^{2} \sim R^{2}\delta (q - |q_{r}|) $. The free energy for a helical worm-like chain according to \eq{fwlc} becomes 
\begin{align}
\frac{F_{wlc}}{k_{B}T} \sim \frac{LL_{P}}{2} q_{r}^{4} R^{2},
\end{align}
which is equivalent to the argument of the Boltzmann distribution [\eq{fsingle}].

\bibliographystyle{apsrev4-1}
\bibliography{fall_et_al}

\end{document}